\documentclass[twocolumn,apjl]{emulateapj} 
\usepackage{times}

\shorttitle{A Common Origin for Quasar EELRs and BLRs} 
\shortauthors{Fu and Stockton}
\journalinfo{Accepted by ApJL}
\submitted{Accepted by ApJL}

\newcommand{\othree}{{[O\,{\sc iii}]}}

\newcommand{\nii}{{[N\,{\sc ii}]}}
\newcommand{\nv}{{N\,{\sc v}}}
\newcommand{\civ}{{C\,{\sc iv}}}
\newcommand{\heii}{{He\,{\sc ii}}}
\newcommand{\eg}{e.g.,}

\begin{document}
\title{A Common Origin for Quasar Extended Emission-Line Regions and
Their Broad-Line Regions}
\author{Hai Fu and Alan Stockton\altaffilmark{2}}
\affil{Institute for Astronomy, University of Hawaii, 2680 Woodlawn
Drive, Honolulu, HI 96822}
\altaffiltext{2}{Also at Cerro Tololo Inter-American Observatory, Casilla 603, La Serena, Chile}

\begin{abstract}
We present a correlation between the presence of luminous extended
emission-line regions (EELRs) and the metallicity of the broad-line
regions (BLRs) of low-redshift quasars.  The result is based on
ground-based \othree\ $\lambda5007$ narrow-band imaging and {\it Hubble
Space Telescope} UV spectra of 12 quasars at $0.20 \leq z \leq 0.45$.
Quasars showing luminous EELRs have low-metallicity BLRs (Z $\lesssim$
0.6 Z$_{\odot}$), while the remaining quasars show typical metal-rich
gas (Z $>$ Z$_{\odot}$). 
Previous studies have shown that EELRs themselves also have low metallicities (Z
$\lesssim$ 0.5 Z$_{\odot}$). 
The correlation between the occurrence of EELRs and the metallicity of
the BLRs, strengthened by the sub-Solar metallicity in both regions,
indicates a common external origin for the gas, almost certainly from
the merger of a gas-rich galaxy.
Our results provide the first {\it direct} observational evidence that
the gas from a merger can indeed be driven down to the immediate
vicinity ($<$ 1 pc) of the central black hole.
\end{abstract}


\section{Introduction}

Massive ionized nebulae having characteristic dimensions of a few
$\times10$ kpc surround roughly half of the quasars that are also
steep-spectrum radio sources at $z < 0.5$ (see \citealt{Sto06} for a
review). These luminous extended emission-line regions (EELRs) typically
show complex filamentary structures that bear no close morphological
relationships either with the host galaxies or with the extended radio
structures \citep{Sto87}, and chaotic kinematics uncoupled with the
stars.  There is accumulating evidence \citep{Fu06,Fu07} that these
EELRs comprise gas that has been swept out by a galactic superwind
resulting from feedback from the quasar (\eg\ \citealt{Di-05}).
However, because the presence of a powerful radio jet seems to be a
necessary (though not sufficient) condition for producing a luminous
EELR, it is likely that the superwind is  produced  by a
large-solid-angle blast wave accompanying the production of the radio
jet \citep{Fu07}, rather than by radiative coupling of the quasar's
luminosity to the gas.

The broad-line regions (BLRs) of quasars comprise material concentrated
within $\sim1$ pc of the central black hole (BH). Because of their
proximity to quasar central engines and the accessibility of their
emission lines, BLRs are the most widely used diagnostic for quasar
abundances.
The major metallicity indicators rely on line flux ratios involving
nitrogen lines, due to the ``secondary" nature of the element
\citep{Pag81}. 
Spectra of the BLRs, combined with photoionization models, show that
most of the quasars are metal rich at all redshifts (Z $>$ Z$_{\odot}$;
\citealt{Ham99,Nag06b}). Since quasars are usually hosted by high-mass
galaxies, which typically have a high metallicity for their interstellar
media (the mass---metallicity correlation; \eg\ \citealt{Tre04}), the
high metallicity of quasars is not unexpected from the standpoint of
normal galactic chemical evolution.

Simulations show that, during a galactic merger, the interstellar gas in
the galaxies rapidly loses angular momentum, resulting in massive gas
concentrations near the center of the merged galaxy (\eg\
\citealt{Bar96}). Hence, a merger could potentially feed the BH and
trigger an active galactic nucleus (AGN) or a quasar. If the current
episode of quasar activity was triggered by a recent merger, and the
EELRs were driven out by a superwind from the central part of the
galaxy, then it is possible that there may be some relation between the
gas in the EELRs and that in the BLRs.
In this Letter we explore this possibility by comparing the BLR
metallicity of quasars associated with luminous EELRs with those that do
not show EELRs. 

\section{The Sample}

\begin{table*}
\footnotesize
\begin{center}
\caption{Quasar Sample
\label{tab:prop}}
\begin{tabular}{clcccccc}
\hline
Designation & \multicolumn{1}{c}{Name} & $z$ & M$_{B,QSO}$ & M$_{R,gal}$ & log(M$_{BH}$/M$_{\odot}$) & \nv/\civ & \nv/\heii 
\\
(1) & \multicolumn{1}{c}{(2)} & (3) & (4) & (5) & (6) & (7) & (8)
\\
\hline
\multicolumn{8}{c}{EELR Quasars} \\
\hline
1104$+$7658 &        3C 249.1 & 0.312 & $-25.2$ & \nodata &  8.96 & $<0.010$                  & $<0.046$ \\ 
1427$+$2632 &   B2 1425$+$26  & 0.366 & $-25.4$ & $-23.2$ &  9.73 & $<0.015$                  & $<0.073$ \\ 
1514$+$3650 &    B2 1512$+$37 & 0.371 & $-24.1$ & $-23.2$ &  9.22 & $0.042^{+0.007}_{-0.005}$ & $0.238^{+0.039}_{-0.031}$ \\ 
1547$+$2052 &       3CR 323.1 & 0.264 & $-23.7$ & $-23.1$ &  9.10 & $<0.012$                  & $<0.056$ \\ 
2137$-$1432 &  PKS 2135$-$14  & 0.200 & $-24.2$ & $-23.2$ &  9.15 & $0.037^{+0.008}_{-0.007}$ & $0.423^{+0.097}_{-0.078}$ \\ 
2254$+$1136 &        4C 11.72 & 0.326 & $-24.9$ & \nodata &  9.15 & $0.033^{+0.083}_{-0.015}$ & $0.190^{+0.477}_{-0.099}$ \\ 
\hline
\multicolumn{8}{c}{non-EELR Quasars} \\
\hline
0005$+$1609 &   PKS 0003$+$15 & 0.450 & $-25.1$ & \nodata &  9.24 & $0.256^{+0.071}_{-0.006}$ & $0.930^{+0.260}_{-0.038}$ \\ 
0755$+$2542 &        OI$-$287 & 0.446 & $-22.9$ & \nodata &  7.47 & $0.426^{+0.249}_{-0.135}$ & \nodata \\ 
1052$+$6125 &        4C 61.20 & 0.422 & $-24.3$ & \nodata &  9.57 & $0.157^{+0.069}_{-0.023}$ & $0.903^{+0.414}_{-0.145}$ \\ 
1153$+$4931 &         LB 2136 & 0.334 & $-22.9$ & $-23.8$ &  8.95 & $0.174^{+0.023}_{-0.011}$ & $0.808^{+0.112}_{-0.055}$ \\ 
1704$+$6044 &          3C 351 & 0.372 & $-25.5$ & $-23.7$ &  9.15 & $0.206^{+0.026}_{-0.021}$ & $0.712^{+0.091}_{-0.086}$ \\ 
2311$+$1008 &   PG 2308$+$098 & 0.433 & $-25.4$ & \nodata &  9.30 & $0.274^{+0.021}_{-0.008}$ & $1.427^{+0.123}_{-0.073}$ \\ 
\hline
\end{tabular}
\end{center}
\vskip 2pt
\small
{\sc Notes. ---}
(1) Quasar J2000 designation, 
(2) common name, 
(3) redshift, 
(4) quasar absolute $B$-band magnitude ($k$-correction applied;
converted from \citealt{Vro06}),
(5) host galaxy absolute $R$-band magnitude (after $k$-correction and passive
evolution correction; following \citealt{Lab06}),
(6) black hole masses estimated from \civ\ FWHM and $\lambda L_{\lambda}$ at 1350 \AA, 
using a formula based on virial method \citep{Lab06},
(7, 8) UV emission line ratios and 1$\sigma$ uncertainty \citep{Kur02,Kur04}.  
\end{table*}

We have compiled a sample of steep-spectrum radio-loud quasars that have
both {\it Hubble Space Telescope} ({\it HST}) Faint Object Spectrograph
(FOS) spectra covering the \nv\ $\lambda1240$ and \civ\ $\lambda1549$
and/or \heii\ $\lambda1640$ lines emitted by the BLRs (hereafter \nv,
\civ\ \& \heii; \citealt{Kur02,Kur04}), and \othree\ $\lambda5007$
narrow-band imaging data to detect or put upper limits on any EELRs
associated with the quasar \citep{Sto87}.
We ended up with 6 objects that show luminous EELRs (the luminosity of
the extended \othree\ emission, $L_{\rm[O\,III]} > 5\times10^{41}$ erg
s$^{-1}$; hereafter the ``EELR quasars''), and 6 ``non-EELR'' quasars
(3$\sigma$ upper limits of $L_{\rm[O\,III]} < 3\times10^{41}$ erg
s$^{-1}$). Here, we have based the EELR luminosities on the ``peak''
luminosities given in Table 1 of \citet{Sto87}, since the upper limits
to the ``total'' luminosities given there necessarily assume an
unrealistically smooth distribution of emission.  The quasar redshifts
range from 0.2 to 0.45.  The radio powers and spectral indices of the two 
subsamples are similar.
\cite{Kur02,Kur04} have given measurements of broad emission lines in
the FOS spectra, from which we calculated the \nv/\civ\ and/or
\nv/\heii\ line ratios and estimated their 1$\sigma$ uncertainty using
the standard error propagation formula.

We have also obtained the absolute $B$-band magnitude (M$_{B,QSO}$) of
the objects from \cite{Vro06}. 
The absolute $R$-band magnitude of the host galaxies (M$_{R,gal}$) is
available for 6 of them, which have been imaged by {\it HST} WFPC2 with
a broad-band filter \citep{Lab06}. 
Using a formula based on the virial theorem \citep{Lab06}, the black
hole masses (M$_{BH}$) were estimated from the \civ\ FWHM and the
continuum luminosity ($\lambda L_{\lambda}$) at 1350 \AA, which are
available from the modeling of the FOS spectra \citep{Kur02,Kur04}. Our
BH mass results are in agreement with those of \citet{Lab06} within a
factor of two.
All of the data tabulated in Table~\ref{tab:prop} have been scaled to a
$\Lambda$CDM cosmology with $H_0$ = 70 km s$^{-1}$ Mpc$^{-1}$,
$\Omega_M=0.3$ and $\Omega_{\Lambda}=0.7$. 

\section{Results}

\begin{figure*}[!t]
\epsscale{0.87}
\plotone{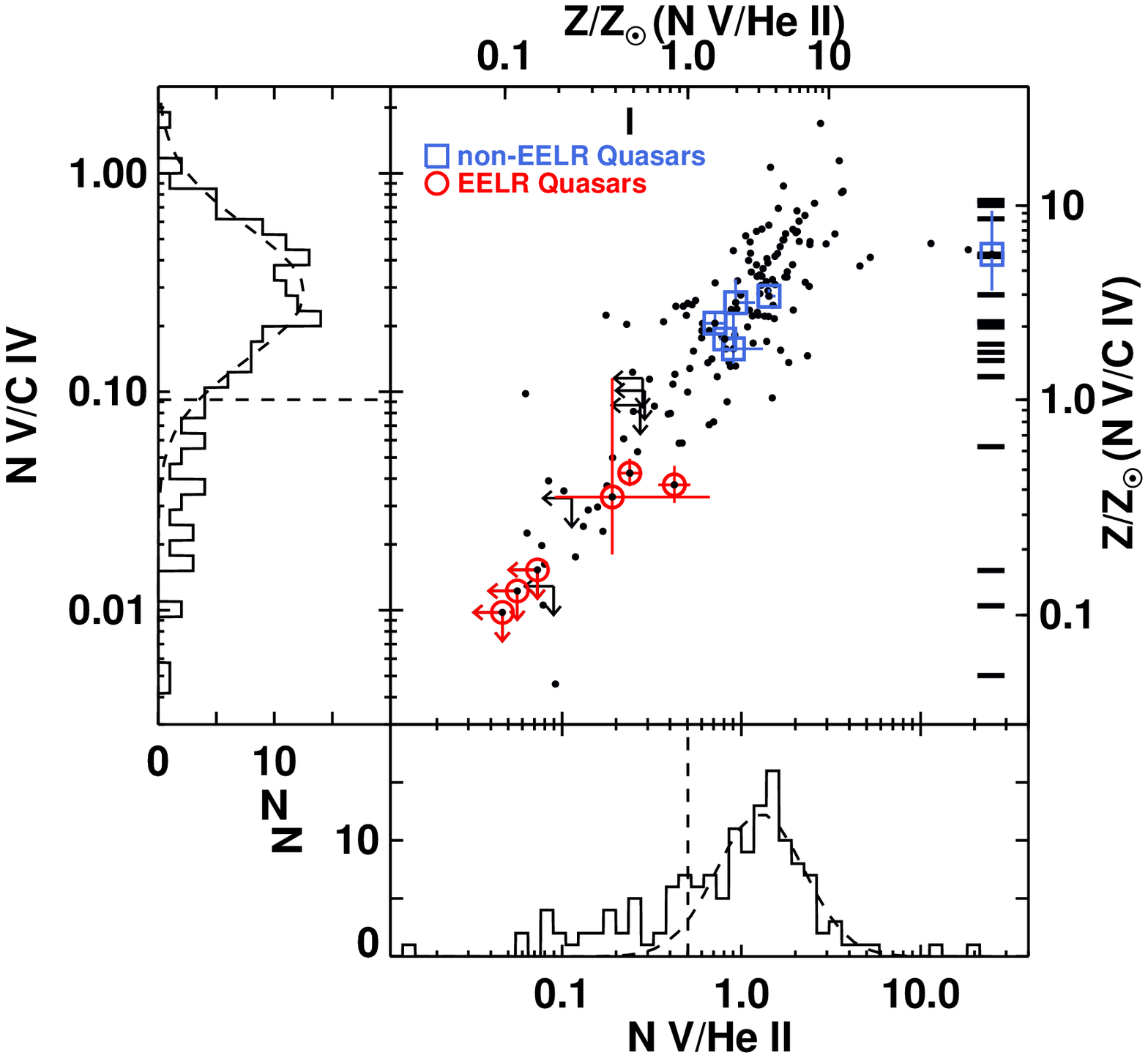}
\caption{
\nv/\civ\ line ratios vs. \nv/\heii\ line ratios. The quasars in the
{\it HST} FOS sample \citep{Kur02,Kur04} with measurements of all three
lines are shown as black points and arrows (3$\sigma$ upper limits) in
the main panel. The bars aligned on the right and upper edges show the
objects with only \nv/\civ\ and only \nv/\heii\ ratios available,
respectively. The EELR quasars are circled in red, and the non-EELR
quasars are blue squares.  The 1$\sigma$ line-ratio errors are also
shown for these objects. The metallicity predicted by the
photoionization model \citep{Nag06b} appears across the right and top
axes. Histograms of \nv/\civ\ and \nv/\heii\ line ratios for all objects with
solid measurements in the entire {\it HST} FOS sample are
shown in the left and bottom panels, respectively.  The dashed lines
mark the Solar metallicity.  Gaussian fits to the histograms are shown
as dashed curves. \label{fig:nv}} 
\end{figure*}

Figure~\ref{fig:nv} shows that the EELR quasars and non-EELR ones are
clearly separated by their broad-line ratios. \nv/\civ\ and \nv/\heii\
flux ratios are predicted to increase substantially with metallicity in
BLRs \citep{Ham02}. The validity of using these two line ratios as
metallicity indicators has been confirmed by comparing results from
other weaker nitrogen lines \citep{Bal03,Dha07}. Therefore, we conclude
that the metallicity of the EELR quasars is systematically {\it lower}
than that of the non-EELR quasars. Specifically, from a calibration of
the line ratios in terms of metallicities \citep{Nag06b}, the
metallicity of the former group ranges from $\sim$0.1 to 0.6
Z$_{\odot}$, compared to 1 to 5 Z$_{\odot}$ for the latter. The Solar
elemental abundances are defined by \citet{And89}.

On the other hand, the two groups look surprisingly similar in terms of
other parameters, such as the redshift range, the quasar luminosity, the
luminosity of the host galaxy and the black hole mass (refer to
Table~\ref{tab:prop}). 

The EELR quasars are obvious outliers with respect to the observed
metallicity---quasar luminosity correlation \citep{Ham99,Nag06b}, or the
purported metallicity---BH mass \citep{War03} and
metallicity---accretion rate \citep{She04} relations ($\dot{M} =
L/L_{edd} \propto 0.398^{M_{B,QSO}}/M_{BH}$). Furthermore, their
low metallicity is also incompatible with the observed tight
mass--metallicity correlation of normal galaxies \citep{Pag81,Tre04}, if
the gas is from the interstellar medium of a galaxy as massive as the
quasar hosts. Like normal AGN, the 4 (out of 6) EELR quasars for which
the host galaxy luminosity were available follow the BH mass---bulge
luminosity relation \citep{Lab06}. Thus, if the co-evolution of galaxies
and their central black holes is indeed responsible for establishing
this correlation, then for black holes of these masses, the accompanying
star formation should have enriched the interstellar media in these
galaxies to super-Solar metallicities. The observed low metallicity of
the gas thus indicates that it originates {\it externally} to the quasar
host galaxies. 

The lower metallicity of the EELR quasars compared to the non-EELR ones
implies some sort of link between gas in the close vicinity of a BH ($<$
1 pc) and the material far out in the galaxy ($>$ 10 kpc).  There is
evidence that the EELRs also have a much lower metallicity when compared
with the typical emission-line gas in an AGN.  The optical line ratio
\nii\ $\lambda6584$/H$\alpha$, when combined with \othree\
$\lambda5007$/H$\beta$, offers a convenient metallicity calibration for
low-density gas photoionized by an AGN. This calibration has been used
in the narrow-line regions of Seyfert 2 galaxies, and it has been shown
to yield consistent metallicity with those extrapolated from nuclear
H\,{\sc II} regions \citep{Sto98}. The same calibration can be used to
infer abundance for EELRs, since the EELRs are also photoionized and
represent a similar density regime.  For 3 of the 6 EELR quasars in our
sample (1104+7658, 1514+3650, and 2254+1136), flux measurements for the
key nitrogen line are available for their EELRs \citep{Bor84,Fu06,Fu07}.
The line ratios of all three EELRs are different from, and on the lower
metallicity side of, most of the AGN narrow-line regions at similar
redshifts.  Specifically, the EELRs have a gas phase metallicity Z
$\sim$ 0.5 Z$_{\odot}$ \citep{Sto02,Fu06,Fu07}, and most AGN narrow-line regions
have Z $>$ Z$_{\odot}$ \citep{Gro06}. 

The correlation between the occurrence of EELRs and the metallicity of
the quasar BLRs, reinforced by the similar metallicity of the EELRs and
the BLRs, suggests a common origin of the two.

\section{Origin of The Gas}

Cooling flows could in principle explain both the external origin and
the sub-Solar metallicity of the emission-line gas of EELR quasars.
However, this scenario in practice seems to have been ruled out by deep
{\it Chandra} X-ray observations of four EELR quasars\footnote{Two of
the four quasars are in our sample.} \citep{Sto06b}, since the hot halo
gas ($T \sim 10^7$ K) from which the warm emission-line gas is suggested
to condense is not detected. Furthermore, a photoionization model
\citep{Sto02} of a representative EELR indicates that the clouds largely
comprise a warm low-density medium, which has a pressure far too low to
be in hydrostatic equilibrium with a hot external medium that would have
a cooling time less than a Hubble time. Therefore, the merger of a
gas-rich galaxy seems to be the most likely explanation for an external
origin of the gas. Indeed, the disturbed morphology of the host galaxies
of at least some EELR quasars (\eg\ 3C\,48, \citealt{Can00};
B2\,1512+37, \citealt{Sto02}) clearly indicates ongoing mergers.

Assuming (1) the BH has built up most of its mass at a much higher
redshift, (2) the current nuclear activity in these EELR quasars is
triggered by a recent merger, and (3) both the EELR and the BLR have
their origins in the interstellar gas of the incoming galaxy, we can put
some constraints on the ``intruder" based on the properties of the
emission-line gas. 
The total ionized mass of a typical luminous EELR is $\sim10^{9-10}$
M$_{\odot}$ (\citealt{Fu06,Fu07}; in comparison, the BLR contains a
negligibly small amount of mass, with estimates ranging from 1 to $10^4$
M$_{\odot}$) and the metal abundance is about 1/2 Z$_{\odot}$ or less.
The intruding galaxy must therefore contain a substantial amount of
metal-poor interstellar gas.  The only types of galaxies we are aware of
that potentially meet these requirements of substantial gas mass
combined with low metallicity are the low surface brightness disk
galaxies (\eg\ \citealt{Hoe00}) and perhaps some late-type spiral
galaxies. 

It is unclear why a merger of a more massive normal spiral, which may
have a similar amount of gas (although a smaller gas fraction and a
higher gas-phase metallicity), would not also produce an EELR.  One
possibility is that the higher metallicity will lower the accretion rate
of material towards the center, because both the higher opacity of the
gas and larger amount of dust will couple the gas more efficiently to
the radiation field of the quasar.  How such a lowered accretion rate
will affect the development of the quasar is not certain, but, if it
delays the formation of the radio jet, then much of the gas may have
time to form stars before the jet is produced.

At the other end of the mass scale, if a merger with a gas-rich dwarf
galaxy (\eg\ a blue compact galaxy or an irregular galaxy) could also
trigger a quasar, then any EELR associated with this quasar would not
have been in our sample, since the total mass of the interstellar gas
would be below the detection limit of the EELR survey of \citet{Sto87}.
However, the BLR would certainly be seen as having a low metallicity due
to the accretion of the metal-poor gas by the BH. Therefore, the lack of
detection of any metal-poor, non-EELR quasars implies that such mergers
are not able to trigger a quasar (they may, however, trigger a
low-luminosity AGN; \citealt{Tan99}), probably because it takes too long
($>$ a few Gyr) to complete such a merger and most of the gas would have
been stripped away by tidal forces and the ram pressure of the halo gas
before the dwarf makes its way to the nucleus \citep{May06}.

Galactic merging has long been suggested to be a major mechanism for
igniting nuclear activity in galaxies, since the interaction can bring 
fresh fuel close to the central BH. However, there exists only indirect 
evidence for the ability of mergers to deliver the gas sufficiently close 
to the nuclei to be accreted by the BHs (\eg\
\citealt{Zir96,San96,Can01,Hae93}), and numerical simulations to
date do not have sufficient dynamic range to explore such small scales.
%
In the EELR quasars, the low metallicity of the gas in the EELR points
to an external origin, most likely from the merger of a gas-rich galaxy.
The correlation in metallicity between the gas at large scales and that
in the BLR then provides the first {\it direct} observational
evidence that the gas from a merger can indeed be driven down to the 
immediate vicinity of the central black hole.

There has been much recent discussion on the relative importance of
``quasar mode'' and ``radio mode'' feedback in controlling galaxy
formation and establishing the bulge-mass---BH-mass correlation.
Quasar-mode feedback is usually envisioned as the radiative coupling of
some of the energy output of a quasar to the surrounding gas, which
expels the gas and quenches further star formation in the forming galaxy
\citep[e.g.,][]{Hop06}.  Radio-mode feedback involves the prevention of
surrounding gas from cooling sufficiently to form stars by the
thermalization of the mechanical energy of radio jets (mostly FR\,I;
e.g., \citealt{Bes06}).  Our results suggest that there may also be a
place for a variant of radio-mode feedback, operating exclusively in
FR\,II radio sources, in which a wide-solid-angle blast wave from the
production of the radio jet impulsively sweeps out a large mass of gas,
in a manner quite similar to that envisioned for quasar-mode feedback.
Because of the peculiar and poorly understood limitation of this mode to
low-metallicity gas and the likely need for a BH that has already
acquired a substantial mass, it might seem that this mechanism may have
limited applicability in the early universe.  Nevertheless, it is not
unreasonable that such a scenario could occur rather frequently during
the formation stage of the most massive galaxies.

\acknowledgments 
This research has been partially supported by NSF grant AST03-07335. 



\end{document}